\documentstyle[aps,preprint,graphicx]{revtex}
\begin{document}
\draft
%\twocolumn[\hsize\textwidth\columnwidth\hsize\csname@twocolumnfalse\endcsname

%\begin{document}
%\draft
\title{Aging Induced Multifractality} 

\author{P. Allegrini$^{1,2}$,
J. Bellazzini$^{3}$, G. Bramanti$^{4}$, M.
Ignaccolo$^{2,5}$, P. Grigolini$^{2,4,6}$ and J. Yang $^{2}$ }

\address{$^{1}$Istituto di Linguistica Computazionale del Consiglio
Nazionale delle Ricerche,\\ Area della Ricerca di Pisa-S. Cataldo, Via
Moruzzi 1, 56124, Ghezzano-Pisa, Italy}

\address{$^{2}$Center for Nonlinear Science, University of North
Texas, P.O. Box 311427, Denton, Texas, 76203-1427}

\address{$^{3}$ Dipartimento di Ingegneria Aerospaziale
dell'Universit\`a di Pisa, Via Caruso, 56100 Pisa, Italy}

\address{$^{4}$Dipartimento di Fisica dell'Universit\`a di Pisa and
INFM, Piazza Torricelli 2, 56127 Pisa, Italy\\}

\address{$^{5}$Center for Nonlinear Science, Texas Woman's University,
P.O. Box 425498, Denton, Texas 76204}

\address{$^{6}$Istituto di Biofisica del Consiglio Nazionale delle
Ricerche,\\ Area della Ricerca di Pisa-S. Cataldo, Via Moruzzi 1,
56124, Ghezzano-Pisa, Italy}

\date{\today} 

\maketitle

\begin{abstract}    
We show that the dynamic approach to L\'{e}vy statistics is
characterized by aging and multifractality, induced by an ultra-slow
transition to anomalous scaling. We argue that these aspects make it a
protoptype of complex systems.
%\end{abstract}

\noindent
\end{abstract}
%\vskip2pc]
%\narrowtext

The dynamic approach to L\'{e}vy statistics is still a poorly
understood problem in spite of several attempts made at deriving it
from intermittency\cite{geisel,shlesingerklafter,west} with several
techniques ranging from Continuous Time Random Walk
(CTRW)\cite{klafter} to the Nakajima-Zwanzig projection
method\cite{allegro,juri}. This letter establishes that the dynamic
approach implies aging, multiscaling and multifractality, thereby
explaining, among other things, why this issue escaped so far a
satisfactory understanding. The aim of this letter is to contribute to
the comprehension of this delicate issue. Before addressing the
problem from a more technical perspective that will be by necessity
less accessible to a general audience, we want to illustrate the
problem, and the solution of it afforded by this letter as well, with
intuitive and qualitative arguments. To make our illustration as clear
as possible, we adopt the same tutorial approach as that used by the
authors working in the field of Cryptography\cite{schneier}, and we
introduce two new archetypal individuals, Bob and Jerry. We hope that
Bob and Jerry might have, with aging induced multifractality, the same
fortune as Alice and Bob with Cryptography. These two individuals aim
at realizing a process of diffusion of L\'{e}vy type, with an
intensity that depends on secrete numbers known only to them. Since
the width of the probability distribution does not depend only on
time, but also on the intensity of the jumps made by the random
walkers at any unit time, it is hard, in principle, to establish the
time at which the diffusion process began, if the intensities of the
jumps are not known. However, as we shall see, this is possible with
Bob's experiment, while it is impossible with Jerry's experiment. In
fact, Bob and Jerry realize L\'{e}vy diffusion in two different ways,
and, as we shall see, Bob, who adopts the dynamic approach, allows us
to predict the exact time at which he started the experiment, while it
is not possible to guess the right starting time in the case of
Jerry's experiment. Bob generates a random sequence of pairs
$\{\tau_{i}, s_{i}\}$, with $i = 0,1,...\infty$. The first number of
each pair, $\tau_{i}$, is randomly drawn from the distribution

\begin{equation}
\label{distribution}
\psi(\tau) =
\frac{(\mu -1)T^{\mu-1}}{(T+ \tau)^{\mu}}.
\end{equation}

The index $\mu$ has to fit the condition $\mu > 2$, which ensures that the mean waiting time $\langle\tau\rangle$,
\begin{equation}
\langle\tau\rangle = \frac{T}{\mu-2} ,
\label{meantime} 
\end{equation} 
is finite. 
Thus the number $T$ controls the intensity of $\langle\tau\rangle$.
The second symbol, $s_{i}$, is a sign, $+$ or $-$, and it is obtained
by tossing a fair coin. Let us imagine that this distribution is used
to generate a diffusion process according to the following
prescription. Bob, who has at his diposal a virtual infinite number of
walkers, creates, for any of them, a sequence $\{\tau_{i},
s_{i}\}$ and makes her travel with velocity $s_{i} W$ for the time
$\tau_{i}$. Then Bob selects a number $r$ belonging to the interval 
$(0, \tau_{0}]$. The space travelled by the walker 
at time $t > \tau_0 -r$ is given by
\begin{equation}
\label{displacement}
\begin{array}{ccc}
 x(t) &=& W[ (\tau_{0} - r) s_{0} + \tau_{1}s_{1}+\cdots \\
&& +\tau_{N-1} s_{N-1} + (\tau_{N} - r^{\prime}) s_{N} ],
\end{array}
 \end{equation}
where
$r^{\prime} \equiv (\tau_{0} + \cdots + \tau_{N}) -r -t$
and $N$ denotes the number of drawings made 
by Bob for his walker within the time interval $[0,t]$.
	At times  
	$t < \tau_0 -r$ ($N=0$) the space is given by
\begin{equation}
x(t)=W(\tau_0 -r -r^{\prime})s_0,
\end{equation}  
	with $r^{\prime}= \tau_0 -r -t$, yielding $x(0)=0$.
Bob keeps secret both the
value of $W$ and the value of $T$. He can try to make his diffusion
process look older by increasing either both or only one of these
secrete numbers. Furthermore, to erase any possible form of aging, he selects
the number $r$ randomly. In fact, this has the effect of ensuring the stationary condition
used by the authors of Refs\cite{geisel,klafter} to realize
L\'{e}vy diffusion, under the form of L\'{e}vy
walk\cite{geisel}, which seems to be more realistic than the flight
prescription\cite{shlesingerklafter,west}.

To appreciate the properties of the L\'{e}vy walk, realized by Bob's
experiment, it is convenient to contrast it with Jerry's
experiment. Also Jerry has at his disposal a virtually infinite number
of random walkers, whose position at $t =0$ is $x = 0$, and he too,
for any of his random walkers, selects an infinite sequence
$\{x_{i}\}$. The numbers $x_{i}$ are drawn from a symmetric
distribution, the positive numbers having the same probability as the
negative numbers. For this reason Jerry does need the coin tossing to
select $s_{i}$. Furthermore, Jerry makes his walker jump at any time
step, by a jump of intensity $|x_{i}|$, in the positive or negative
direction according to the sign of $x_{i}$. To be more precise, let us
say that the distribution used by Jerry is $\Pi(x)$, defined through
its Fourier transform, 
\begin{equation} 
\hat\Pi(k) = \exp(-b|k|^{\mu -1}),
\label{levy} 
\end{equation} 
and only Jerry knows the secrete value of $b$. The distribution of
numbers used by Jerry is the well known L\'{e}vy
distribution\cite{shlesinger}: a stable distribution yielding a
diffusion process, with the probability distribution $p_{L}(x,t)$,
whose Fourier transform is

\begin{equation} 
\hat p_{L}(k,t) = \exp(-b|k|^{\mu -1} t).
\label{levydiffusion} 
\end{equation}
The time $t$ is the number of drawings, but it is so large as to be
virtually indistiguishable from a continuous number. It is clear that
the observation of Jerry's diffusion process does not allow any
observer to establish when he began his experiments. We assume that
the observer, which might be a third archetypal individual, does not
know the time at which Jerry began his diffusion. By means of the
experimental observation he/she can only establish $bt$, and since $b$
is not known to him/her, he/she cannot determine the value of $t$. In
other words, a broad distribution can be the consequence of Jerry
starting his diffusion process at a very early time, but it can also
be the consequence of a late beginning with much more intense jumps.

It is not so with Bob's diffusion process. Let us see why. Let us
consider, as in the case of Jerry's experiment, a time $t$ very
large. In the case of interest here, $\mu > 2$, the mean waiting time,
Eq. (\ref{meantime}), is finite. Thus, the number of random drawings
and coin tossings is 
%given
very well approximated 
by $N = t/\langle\tau\rangle$. Using the
Generalized Central Limit Theorem (GCLT) \cite{gnedenko} we predict
that Jerry's experiment yields the same statistics as Bob's
experiment, L\'{e}vy statistics. However, this important theorem does
not afford any clear indication about the time necessary to realize
this statistics. The predictions of the GCLT theorems are
realized\cite{gianmarco} by the following expression for $p(x,t)$:

\begin{equation} p(x,t) = K(t) p_{T}(x,t)
\theta (Wt-|x|) + \frac{1}{2}\delta (|x|-Wt) I_{p}(t),
\label{truncated} 
\end{equation} 
where $\theta$ denotes the Heaviside step function. We also note that
\begin{equation}
\label{correlationfunction} 
\lim_{t \rightarrow \infty} I_{p}(t) = \Phi_{\xi}(t) = \left(\frac{T}{T+t}\right)^{\mu-2},
\end{equation} 
$p_{T}(x,t)$ is a distribution that for $t \rightarrow \infty$ becomes
identical to the anti-Fourier transform of Eq. (\ref{levydiffusion}),
and $K(t)$ is a time-dependent factor ensuring the normalization of
the distribution $p(x,t)$.
 
We observe that at any time $t$, no matter how arbitrarily large, it
is possible to find a significant number of Bolb's walkers, with
$\tau_{0} - r > t$, namely, walkers for which Bob has not yet drawn the
second pair of stochastic numbers. This probability is expressed
by\cite{renewal}

\begin{equation}
\label{renewaltheory}\Phi_{\xi}(t) =
%\frac{1}{\tau_{W}}\int_{t}^{\infty}(t^{\prime} - t) \psi(t^{\prime})
%dt^{\prime}.
\frac{1}{\langle t \rangle}\int_{t}^{\infty}(t^{\prime} - t) \psi(t^{\prime})
dt^{\prime}.
\end{equation}
It is interesting to notice that, due to the fact that Bob decides the
motion direction by tossing a coin, the function $\Phi_{\xi}(t)$ is
the correlation function of the variable $\xi(t)$, namely, the
fluctuating velocity created by Bob's experiment. The number of
walkers contributing to the propagation front is slightly
larger. However, it is straightforward to prove with arguments similar
to those used by Zumofen and Klafter\cite{klafter} that in the
asymptotic time limit $I_{p}(t)$ becomes identical to $\Phi_{\xi}(t)$,
thereby accounting for the former of the two equalities of Eq.
(\ref{correlationfunction}). The latter is easily accounted for by
plugging $\psi(\tau)$ of Eq. (\ref{distribution}) into Eq.
(\ref{renewaltheory}).

On the basis of these arguments we reach the conclusion that in the 
asymptotic time limit Eq. (\ref{truncated}) becomes identical to
(see \cite{allegro} for earlier derivation)

\begin{equation} p(x,t) = p_{L}(x,t)
\theta (Wt-|x|) + \frac{1}{2}\delta (|x|-Wt) \Phi_{\xi}(t).
\label{truncated2} 
\end{equation}
This equation, although valid only in the asymptotic time limit, is
very convenient for the theoretical arguments of this letter. First
of all, it allows us to determine the age of the diffusion experiment
created by Bob, even if Bob adopts the stationary condition\cite{geisel,klafter} and keeps secret the values of $W$ and $T$. To
do so, we measure the distance of one ballistic peak from the other,
the diffusion coefficient $b$ of $p_{L}(x,t)$ of
Eq. (\ref{levydiffusion}) and the intensity of the two ballistic
peaks. All these three quantities can be expressed in terms of the
unknown quantities, $t$, $W$ and $T$. The distance between the two
peaks is: $2 Wt$; the diffusion coefficient is given by: $b = W
(TW)^{\mu-2} sin[(\pi (\mu-2)/2] \Gamma(3-\mu))$ \cite{mario} and the
peak intensity by Eq. (\ref{correlationfunction}). The age of Bob's
experiment can be revealed by means of experimental observation,
thanks to the breakdown of homogeneity (multi-scaling), caused by the
ultra-slow relaxation of $\Phi_{\xi}(t)$. At the intuitive level of
this first part of the letter, we can say that aging causes
inhomogeneity.

Let us now move to a more technical level of description. We share the
vision of Khinchin\cite{khinchin} about the close connection between
ordinary statistical mechanics and the ordinary Central Limit Theorem. 
There is, on the other hand, a close connection between
statistical equilibrium and scaling of a diffusion process. The latter
property is expressed by

\begin{equation} 
p(x,t) = \frac{1}{t^{\delta}}
F(\frac{x}{t^{\delta}}). 
\label{scaling} 
\end{equation}
In fact, this property means that the probability density at 
different times can be expressed in terms of the same time independent
property $F(y)$. The case of ordinary statistical mechanics 
corresponds to $\delta = 1/2$ with $F(y)$ being a Gaussian function 
of $y$. The case under study in this paper is not ordinary because
$F(y)$ is a L\'{e}vy distribution and the scaling parameter $\delta$ 
is given by

\begin{equation} \delta = \frac{1}{\mu -1}. 
\label{levyscaling} 
\end{equation}

This thermodynamic condition, however, is the asymptotic limit of a
very slow transition, driven by the correlation function
$\Phi_{\xi}(t)$ of Eq. (\ref{correlationfunction}), whose lifetime, in
the case here under study, with $2 < \mu < 3$, is infinite. For a
further discussion of this extremely slow transition from dynamics to
thermodynamics, the reader can consult also Ref. \cite{massi}.
Although this process of transition to thermodynamics, or aging, is
extremely slow, with virtually an infinite lifetime, it is not easy to
detect by mean of the current techniques of analysis of time series,
which rest on the observation of a single sequence (see for instance
Ref. \cite{fractals}). Here we show that it turns out to be difficult
even with the most advanced method of scaling detection, the method of
Diffusion Entropy (DE) \cite{fractals}. The first step of all these
techniques, including the DE method, is based on deriving the random
walkers required by Bob's experiments from a single sequence. We use
Bob's algorithm to create a single, virtually infinite, trajectory.
Then we span infinitely many portions of this sequence moving along
the sequence a window of width $t$. The selected walking trajectories
are shifted in such a way as to make them start at $x = 0$ at $t = 0$.
These trajectories tends to depart the ones from the others, as an
effect of their partially random character, and this spreading is
described by $p(x,t)$, evaluated numerically with a proper partition
of the $x$-axis. At this stage, rather than evaluating the variance,
which leads to a wrong scaling in the Non-Gaussian case\cite{nicola},
we evaluate the Shannon entropy

\begin{equation}
\label{shannonentropy}
S(t) = -\int_{-\infty}^{+\infty} p(x,t) \log(p(x,t)).
\end{equation}
In the case when the scaling condition of Eq. (\ref{scaling}) applies,
by plugging Eq. (\ref{scaling}) into Eq. (\ref{shannonentropy}) we get

\begin{equation}
\label{shannonentyropyyieldingscaling}
S(t) = A + \delta \log(t),
\end{equation}
with $A$ being a constant whose explicit expression is of no interest
here. Fig. 1 proves that the DE method is successful in detecting the
predominant asymptotic scaling of Eq. (\ref{levyscaling}). This is a
flattering result, since the methods based on the variance measurement
cannot reveal this L\'{e}vy scaling. However, it is not easy to relate
Fig. 1 to the slow transition process described by
Eq. (\ref{truncated2}) \cite{gianmarco}.

%%%%%%%%%%%%%%%%%%%%%%%%%%%%%%%%%%
%
% figure 1 sould be around here
%
%%%%%%%%%%%%%%%%%%%%%%%%%%%%%%%%%

We can go beyond the limitations of the DE method of analysis
detecting the multifractal properties generated by the aging process
itself. To reveal the emergence of these multifractal properties we
rest on Eq. (\ref{truncated2}) and on a procedure reminiscent of that
of Nakao\cite{nakao}. This author proved in fact that a truncated
L\'{e}vy process yield bifractal properties. It has to be pointed out,
however, that there is a deep difference between the case considered
by Nakao and the dynamic truncation of this letter. The effect
observed by Nakao has to do with the ultraslow convergence to Gaussian
statistics discussed some years ago by Mantegna and
Stanley\cite{mantegnino}, observed independently by the authors of
Ref.\cite{granfica}. The effect here discussed is instead an ultraslow
convergence to L\'{e}vy statistics. Moreover, as we have seen, our
transition process has an infinite lifetime, while the lifetime of
that of Refs.\cite{mantegnino,granfica}, although impressively large,
is finite.

To study the multifractal properties associated to the dynamic
approach to L\'{e}vy processes, we follow the prescriptions of
Refs.\cite{benzi,paladin}. We study theoretically and numerically the
fractional moment $\langle|x|^{q}\rangle$, which is expected to yield

\begin{equation}
\label{fractionalmoment}
\langle|x|^{q}\rangle = \int_{\infty}^{+\infty} p(x,t)|x|^{q} dx \approx t^{\xi_{q}} .
\end{equation}
The power index $\xi_{q}$ plays a critical role. According to the
theory of Refs. \cite{benzi,paladin}, $\xi_{q}$ as a function of $q$,
would be a straight line if the monofractal condition applied, while
its deviation from a straight line signals the occurrence of
multifractal properties. Using the theoretical prediction of
Eq. (\ref{truncated2}) it is straigforward to predict that

 \begin{equation}	
\xi_{q} = \delta q,
\label{smallq}	
\end{equation}
for $q < \mu -1$, and 

\begin{equation} 
\label{largeq} 
\xi_{q} = q - \mu + 2, 
\end{equation} 
for $q > \mu -1$. These theoretical predictions are very
satisfactorily supported by the numerical results, as proved by
Fig. 2.

%%%%%%%%%%%%%%%%%%%%%%%%%%%%%%%%%%%%%%%%%%%%%%%%%%%%%%
%
%  figure 2 should go around here
%
%%%%%%%%%%%%%%%%%%%%%%%%%%%%%%%%%%%%%%%%%%%%%%%%%%%%%%%%%%

In conclusion, we define \emph{aging} the process of transition from
dynamics to thermodynamics when this transition has an infinite
lifetime\cite{note1,note2}. If it had a finite lifetime, it would be
possible to make an observation at times so large as to make the
scaling condition of Eq. (\ref{scaling}) virtually exact. This
condition, in turn, plugged into Eq. (\ref{fractionalmoment}), would
make Eq. (\ref{smallq}) valid for all $q$'s, and not not only for the
small ones. This means that \emph{no aging } is equivalent to the
condition of monofractal statistics. If, on the contrary, aging
exists, the diffusion process becomes bifractal. This is what we mean
by \emph{aging induced multifractality}. We propose aging as a
paradigm for the living state of matter. This has to do with the widely
accepted conviction (see, for instance, Ref.\cite{lifeandcomplexity})
that life is a balance of order and randomness. This paper shows that
aging, as we do mean it, is an attractive expression of this balance.

\begin{figure}[!b]
\begin{center}
\vspace{-1.1cm}
\includegraphics[angle=270,scale=.66]{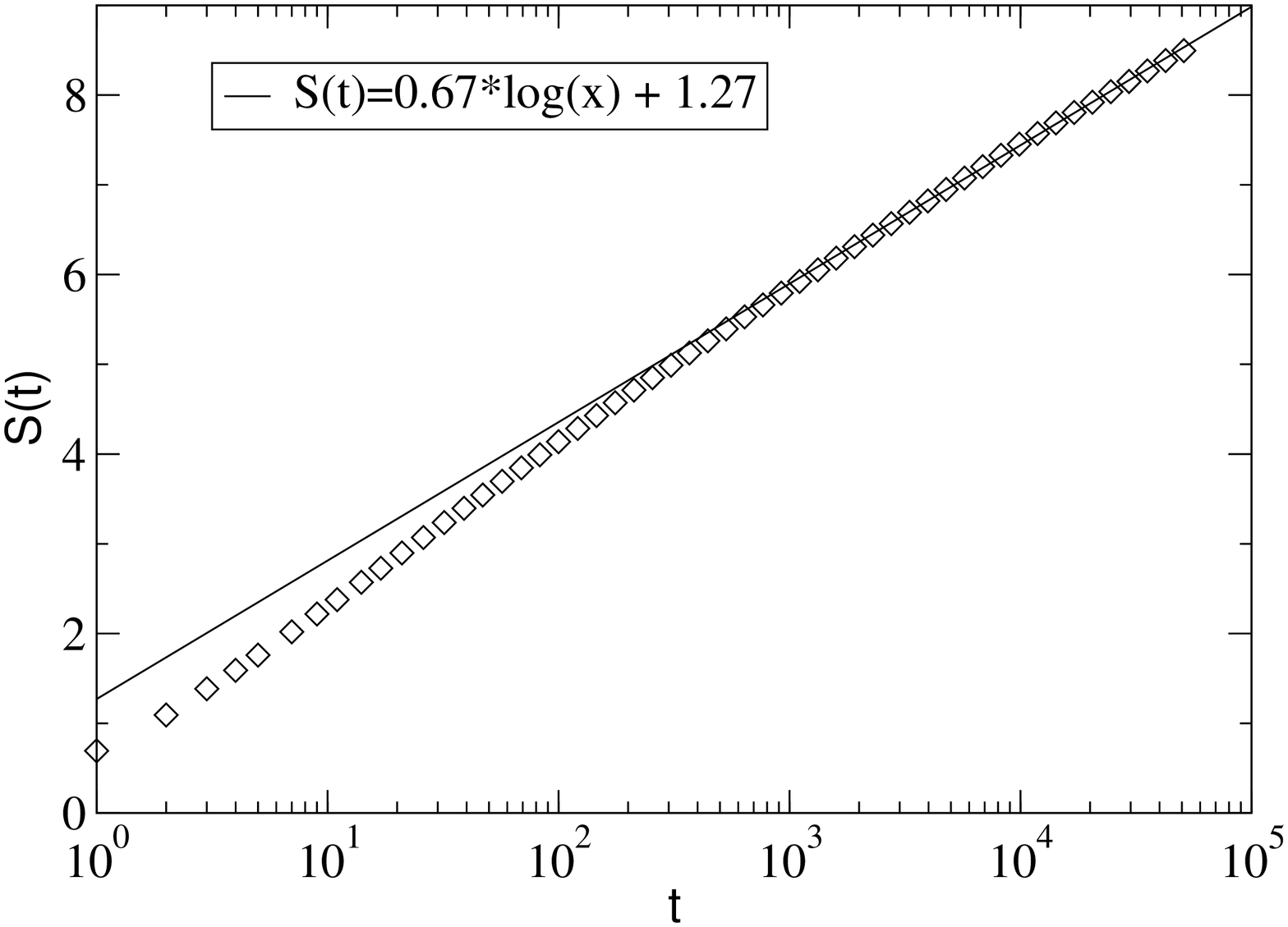}

\label{demu25}
\caption{Diffusion Entropy as a function of time t. Dots: numerical
evaluation of $S(t)$ for Bob's experiment with $\mu=2.5$, $T\approx
1.0$. Bob's walkers are derived from a single trajectory of total
lenght 25,918,673 time units (see the text for details). The solid
line represents a best fit, in the asymptotic regime, of the
theoretical prediction $S(t)=2/3 \log(t) + const$.}
\end{center}
\end{figure}

\begin{figure}[!b]
\begin{center}
\vspace{-1.1cm}
\includegraphics[angle=270,scale=.66]{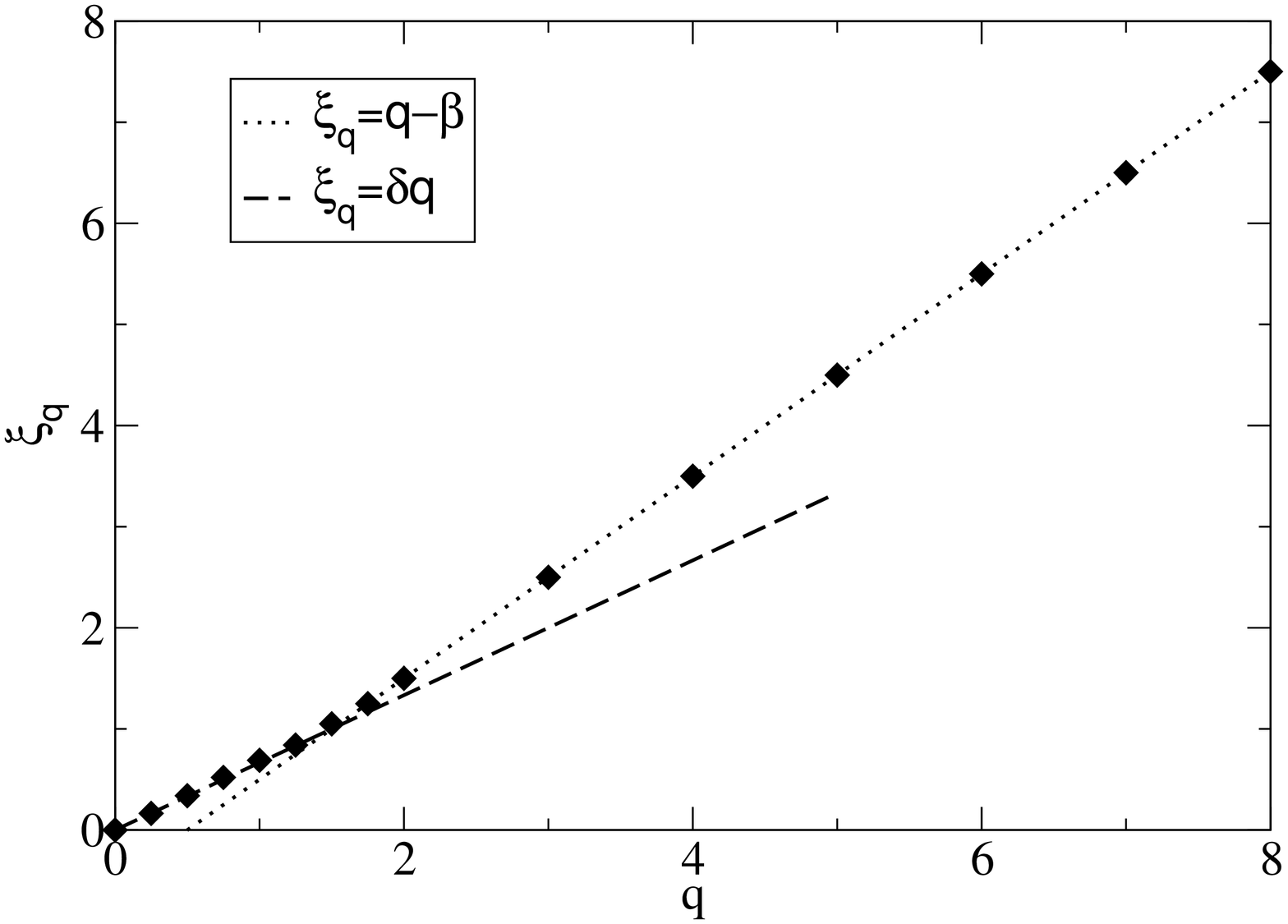}
\label{csiq}
\caption{Multifractal index $\xi_q$ as a function of q. Dots:
numerical evaluation of $\xi_{q}$ for Bob's experiment (dynamical
truncated L\'evy process) using the same data as those of Fig.1.
Dashed line: $\xi_{q}=\delta q$, with $\delta=2/3$ according to
(\ref{levyscaling}); dotted line: $\xi_{q}=q-\beta$, with
$\beta\equiv\mu-2=0.5$.}
\end{center}
\end{figure}


\begin{references}
\bibitem{note2} We do not claim that our definition of aging
corresponds to the spin-glass property with the same name (see
Ref. \cite{note1} for a sample of papers on this literature). We do
not rule out, however, the possibility of an interesting connection.
\end{references}
\end{document}